\documentclass[preprint,12pt]{elsarticle}

\usepackage{amssymb}
\usepackage{url}
\usepackage{epsfig}

\usepackage{lineno}

\newcommand{\prl}{Physical Review Letters}

\newcommand{\mnras}{MNRAS}

\journal{Nuclear Instruments and Methods in Physics Research A}

\begin{document}

\begin{frontmatter}



\title{GWitchHunters: Machine Learning and citizen science to improve the performance of Gravitational Wave detector}


\author[unipi,infnpi]{Massimiliano Razzano\corref{cor1}\fnref{corfootnote}}
\fntext[corfootnote]{on behalf of the REINFORCE Consortium}
\cortext[cor1]{Corresponding author}
\ead{massimiliano.razzano@unipi.it}
\author[unipi,infnpi]{Francesco Di Renzo}
\author[unipi,infnpi]{Francesco Fidecaro}
\author[ego]{Gary Hemming}
\author[ego]{Stavros Katsanevas}

\address[unipi]{Department of Physics, University of Pisa, Largo B. Pontecorvo 3, Pisa, I-56127}
\address[infnpi]{INFN Section of Pisa,  Largo B. Pontecorvo 3, Pisa, I-56127}

\address[ego]{European Gravitational Observatory (EGO),Via E. Amaldi, 5, Cascina,I-56021}

\begin{abstract}
The Gravitational waves have opened a new window on the Universe and paved the way to a new era of multimessenger observations of cosmic sources. Second-generation ground-based detectors such as Advanced LIGO and Advanced Virgo have been extremely successful in detecting gravitational wave signals from coalescence of black holes and/or neutron stars. However, in order to reach the required sensitivities, the background noise must be investigated and removed. In particular, transient noise events called “glitches” can affect data quality and mimic real astrophysical signals, and it is therefore of paramount importance to characterize them and find their origin, a task that will support the activities of detector characterization of Virgo and other interferometers. Machine learning is one of the most promising approaches to characterize and remove noise glitches in real time, thus improving the sensitivity of interferometers. A key input to the preparation of a training dataset for these machine learning algorithms can originate from citizen science initiatives, where volunteers contribute to classify and analyze signals collected by detectors. We will present GWitchHunters, a new citizen science project focused on the study of gravitational wave noise, that has been developed within the REINFORCE project (a "Science With And For Society" project funded under the EU's H2020 program). We will present the project, its development and the key tasks that citizens are participating in, as well as its impact on the study of noise in the Advanced Virgo detector.
\end{abstract}



\begin{keyword}
gravitational waves \sep machine learning \sep citizen science

\PACS 04.20.--q \sep 04.30.Tv \sep 

\MSC 83C35 \sep 

\end{keyword}

\end{frontmatter}


\section{Introduction}
\label{}
Gravitational wave physics is opening an entire new window on the Universe. Since their discovery in 2015 \cite{2016PhRvL.116f1102A}, the Advanced LIGO \cite{2015CQGra..32g4001L} and Advanced Virgo \cite{2015CQGra..32b4001A} detectors have carried on three observing runs (O1, O2, O3) and unveiled 90 signals produced by the coalescence of compact objects, mostly binary black hole with a small fraction of neutron star and/or black hole binaries \cite{2021arXiv211103606T}.\\ 
Advanced LIGO and Virgo are second-generation laser interferometers with Fabry-Perot cavities hosted in km perpendicular arms, that are capable of detecting the tiny deformations induced in the fabric of spacetime by the passage of gravitational waves. In order to improve the sensitivity of the detectors, there is a continuous effort to reduce the background noise due to local disturbances. In particular, at low frequencies the noise is dominated by seismic and Newtonian noise, while at mid frequencies the main component is related to the thermal noise and at high frequencies the noise is mostly related to quantum effects. The activity of detector characterization and noise hunting in gravitational wave detectors is focused on the investigation of stationary and non stationary noise sources. In particular, non stationary transient noise events called \emph{glitches} are of particular importance in the noise studies. In fact, glitches can affect data quality and stability and mimic real astrophysical signals, thus reducing the effective duty cycle of interferometers. The classification and characterization of glitches is therefore key to understand the origin of noise in detector. However, glitches have complex temporal signatures, that make difficult to classify them using standard methods. Various works have shown that Machine Learning methods can be promising for the classification of glitches \cite{2017arXiv171107468G,2015CQGra..32u5012P}. In particular, images built from the time-frequency spectrograms of glitches are very effective in showing the complex morphology of glitches and can be easily given in input to machine learning algorithms, including deep convolutional neural networks \cite{2018CQGra..35i5016R}.
A possible approach to this problem is based on supervised learning, that requires large number of labeled glitch samples, that could be produced by dedicated citizen science initiatives, where volunteers look at images and clssify them. A successful example of this method is provided by \emph{GravitySpy}\footnote{\url{http://https://gravityspy.org/}}, a citizen science project focused on the classification of glitches in LIGO and Virgo\cite{2017CQGra..34f4003Z}. Here we present \emph{GWitchHunters}\footnote{\url{https://www.zooniverse.org/projects/reinforce/gwitchhunters}}, a new citizen science project complementary to GravitySpy and aimed at improving sensitivity of gravitational wave detectors combining citizen science and machine learning.
\section{The REINFORCE Project}
\label{}
GwitchHunters has been developed within the \emph{Research Infrastructures FOR Citizens in Europe} (REINFORCE) project\footnote{\url{https://www.reinforceeu.eu/}}. REINFORCE is a Research \& Innovation Project, supported by the EU H2020 SWAFS “Science with and for Society” work programme and aimed at creating a series of cutting-edge citizen science projects on Frontier Physics research, with the goal of engaging $>$100,000 citizens. REINFORCE is based on four citizen science \emph{demonstrators} focused Gravitational Waves (GWitchHunters), Astrophysical neutrinos (Deep Sea Explorers), High Energy Physics (New Particle Search at CERN) and muon-based tomography (Cosmic Muon Images). All demonstrators are hosted on Zooniverse \cite{2008MNRAS.389.1179L}, the world leading platform for citizen science projects. 
\section{Overview of GWitchHunters}
\label{}
GwitchHunters has been officially launched on Zooniverse in November 2021 after a dedicated review phase and offers to citizens a set of different tasks of increasing difficulty. Data are presented as spectrograms and come from the Virgo O3 run. A \emph{Playground} task is specifically devoted to learning the basics of glitch morphology and its classification. Three other levels offer (1) the possibility to classify glitches among a larger set of classes, (2) localize the glitches in the time-frequency space, and (3) compare the spectrogram in the main channel of Virgo with that produced by auxiliary sensors. This last task is particularly innovative, since it offer the possibility of linking the glitches observed in the main channel to local disturbancies in the detector, thus suggesting a possible hint to the origin of each glitch. These tasks can be carried both on a personal computer and on mobile devices.
The project also features a set of tutorials and examples to teach the volunteers how to perform the different tasks, as well as a "Field Guide" containing information on the Advanced Virgo detector, the various glitch classes and the auxiliary channels used in the project.
\begin{figure}[t]
\centering
\includegraphics[width=0.9\columnwidth]{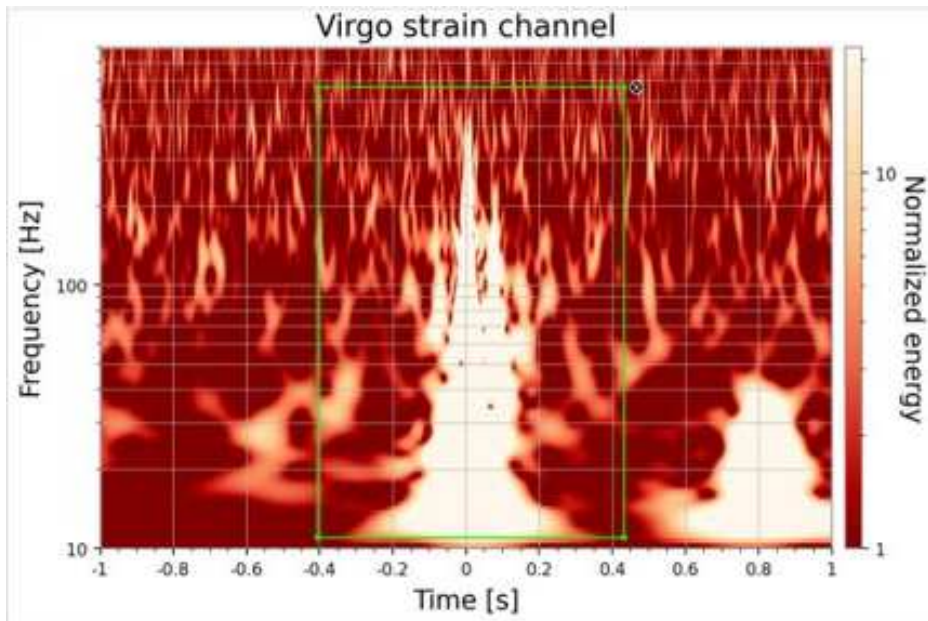}
\caption{Example of a GWitchHunters spectrogram showing two glitches, as well as the rectagle drawn by citizens to locate them}
\label{fig:glitch}
\end{figure}
\section{First Results and Conclusions}
\label{}
Since its official launch in November 2021, $\sim$2800 volunteers have subscribed to the project, although another significant amount have contributed without officially registering. This collective effort has produced more than $\sim$400000 classifications of $\sim$ 40000 data samples so far. In order to promote the project and engage citizens, the REINFORCE consortium has organized many initiatives, including workshops, press activities, online challenges\footnote{e.g. \url{https://www.reinforceeu.eu/winter-challenge-2022}} and training school\footnote{e.g.  \url{https://reinforce.ea.gr/international-training-course/}}. A monitoring of the project website has been carried on, showing that these initiatives successfully attracted more volunteers to GWitchHunters, leading to peaks of $\sim$5000 classifications per day.
The results of the volunteers analysis are used for training a machine learning algorithm that automatically analyze the glitch data. In particular, we focused on a 2D convolutional neural network architecture, that has been also tested on simulations \cite{2018CQGra..35i5016R,2020cuoco} reaching an accuracy greater than 99\%. These first tests show how the GWitchHunters project could be successfully used to join citizen science and machine learning with the goal of contributing to increase the sensitivity of gravitational wave detectors.
\section*{Acknowledgements}
REINFORCE has received funding from the European Union's Horizon 2020 research and innovation program, under Grant Agreement no. 872859.



\end{document}